\documentclass[strcuabstract,twocolumn]{aa}

\usepackage[]{natbib,amsmath,amssymb}
\bibpunct{(}{)}{;}{a}{}{,}
\usepackage{amsmath}
\usepackage{amssymb}
\usepackage{graphicx}
\usepackage{epsfig}

\newcommand{\bea}{\begin{eqnarray}}
\newcommand{\eea}{\end{eqnarray}}
\newcommand{\be}{\begin{equation}}
\newcommand{\ee}{\end{equation}}
\newcommand{\rund}[1]{\left(#1\right)}
\newcommand{\vc}[1]{\mbox{\boldmath $#1$}}
\renewcommand{\d}{{\rm d}}
\newcommand{\dc}{\partial}
\newcommand{\eck}[1]{\left[ #1 \right]}

\newcommand{\abs}[1]{\left| #1 \right|}
\def\eps{{\epsilon}}

\def\elabel#1{\label{eq:#1}}
\begin{document}

  \title{Estimate of dark halo ellipticity by lensing flexion}

   \author{Xinzhong Er\inst{1,2} and
     Peter Schneider \inst{1}
   }
   \offprints{xer}
   \institute{Argelander-Institut f\"ur Astronomie, Universit\"at Bonn,
              Auf dem H\"ugel 71, D-53121 Bonn, Germany\\
          \email{xer, peter@astro.uni-bonn.de}
      \and
      International Max Planck Research School (IMPRS)
      for Astronomy and Astrophysics,
      Auf dem H\"ugel 69, D-53121 Bonn, Germany}

   \date{Received ; accepted}

   \abstract{}
        {The predictions of the ellipticity of the dark matter halos
          from models of structure formation are notoriously difficult
          to test with observations. A direct measurement would give
          important constraints on the formation of galaxies, and its
          effect on the dark matter distribution in their halos. Here
          we show that galaxy-galaxy flexion provides a direct and potentially
        powerful method
          for determining the ellipticity of (an ensemble of)
          elliptical lenses.}
        {We decompose the spin-1 flexion into a radial and
        a tangential
        component. Using the ratio of tangential-to-radial flexion,
        which is independent of the radial mass profile,
        the mass ellipticity can be estimated.}
        {An estimator for the ellipticity of the mass distribution is
        derived  
	and tested with simulations. We show that the estimator is
        slightly biased. We quantify this bias, and provide a
        method to reduce it. Furthermore, a parametric fitting of the
        flexion ratio and
        orientation provides another estimate for the dark halo
        ellipticity, which is more accurate for individual lenses
        Overall, galaxy-galaxy flexion appears as a powerful tool for
        constraining the ellipticity of mass distributions.}
        {}

\keywords{cosmology -- gravitational lensing -- flexion -- dark matter
  -- galaxy halo
}
   \titlerunning{Mass ellipticity from flexion}

   \maketitle

\section{Introduction}
The structure of cluster and galaxy halos predicted by N-body simulations of
Cold Dark Matter show several important features, e.g. the universal radial
density profile \citep{1996ApJ...462..563N,1997ApJ...490..493N}, and the
highly non-spherical structure fitted well by a triaxial density profile
\citep{2002ApJ...574..538J,2009ApJ...703L..67L}. These features are related to
the nature of dark matter as well as to the formation of galaxies and clusters
\citep{2007ApJ...671.1135K,2008ApJ...681.1076D,2010MNRAS.404.1137B}, which
suggest that accurate estimates of the halo properties can provide several
constraints on cosmology.  In particular, numerical simulations with different
assumptions predict halos with different shapes, e.g., simulations with
non-interacting cold dark matter predict that halos are triaxial prolate
ellipsoids \citep{2006MNRAS.367.1781A}. Moreover, if there exists a
significant difference between the shape of a galaxy and the shape of its
total mass distribution, this would provide additional strong evidence for the
existence of dark matter \citep{2010arXiv1007.4815S}. In other words, it
allows us to test how reliable the light
is as a tracer of the matter distribution.

Gravitational lensing provides a powerful tool for studying the mass
distribution of clusters of galaxies as well as galaxy halos
\citep[see][for reviews on weak lensing]{2001PhR...340..291B,
  2003ARA&A..41..645R, 2006glsw.book.....S, 2008PhR...462...67M}.
This is because gravitational lensing probes the matter distribution
regardless of whether it is luminous or dark. The weak lensing
technique has been used for cluster mass reconstructions
\citep[e.g.][]{2006ApJ...648L.109C,2008ApJ...687..959B}, and for the
ellipticity of the dark matter distribution
\citep{2009MNRAS.393.1235C,2009arXiv0912.4260D,2010MNRAS.tmp..941H}.
A method to determine galaxy halo ellipticity by stacking galaxies was
proposed in \citet{2000ApJ...538L.113N} and has been used to determine
the ellipticities of both cluster- or galaxy-size halos
\citep{2009ApJ...695.1446E, 2006MNRAS.370.1008M}. A mean ellipticity
of 0.46 is found from a sample of 25 clusters using Subaru data
\citep{2010arXiv1004.4214O}.

Flexion has been recently studied as the derivative of the shear, and
responds to small-scale variations in the gravitational potential
\citep{2002ApJ...564...65G,2005ApJ...619..741G,2006MNRAS.365..414B}.
Different techniques were developed to measure flexion \citep{
  2006ApJ...645...17I,2007ApJ...660..995O,2007MNRAS.380..229M,
  2008A&A...485..363S}. It was noted that flexion can contribute
to studies of the dark matter halos in galaxies and clusters,
especially for the detection of mass substructure
\citep{2009arXiv0909.5133B,2010arXiv1008.3088E}.  Flexion has been
implemented on small sets of observational data.
\citet{2008ApJ...680....1O} performed flexion measurement on data
from the Subaru telescope to detect substructure in Abell 1689.
\citet{2007ApJ...666...51L} analyzed images taken by the HST Advanced
Camera for Surveys, from which a preliminary galaxy-galaxy flexion
signal has been detected.  \citet{2009MNRAS.400.1132H} studied 
galaxy halo ellipticity using flexion. It is found
that the constrains from flexion are comparable to, or even tighter than
those from shear. Moreover, since flexion drops off faster than the
shear as one goes away from the center of the lens, multiple
deflections from the three-dimensional mass distribution between us
and distant source galaxies \citep{2010MNRAS.tmp..941H} may not be
significant for flexion.

In this paper, we present a new approach to estimate dark matter halo
ellipticity using galaxy-galaxy lensing flexion.  The spin-1 flexion,
a vector, is decomposed into its two components, the radial and
tangential flexion. The tangential-to-radial flexion ratio yields an
estimate of dark halo ellipticity.  We need to assume that the center
of the galaxy or the cluster be known. Whereas for galaxies this may
be a lesser problem, the mass centroid of clusters is sometimes
difficult to determine. We study the effect of noise coming from an
intrinsic flexion.  In Sect. 3, we perform a series of numerical tests
of our estimate and study how our results are biased by the
distribution of background sources, the centroid offsets and intrinsic
flexion. We discuss our results in Sect. 4.

\section{\label{Sc:2}Basic formalism}

The full formalism described here can be found in
\citet{2006MNRAS.365..414B,2008A&A...485..363S}.
Weak lensing shear and flexion are conveniently described
using a complex notation. We adopt the thin lens approximation,
assuming that the lensing mass distribution is projected onto the lens
plane. The dimensionless projected mass density can be written as
$\kappa(\vc \theta)= {\Sigma(\vc\theta) / \Sigma_{\rm cr}}$,
where $\vc\theta$ is the angular position, $\Sigma (\vc\theta)$ is the
projected mass density, and $\Sigma_{\rm cr}$ is the critical surface
mass density.

The first-order image distortion by gravitational lensing is described
by the shear
$\gamma$, which transforms a round source into an elliptical image. The
higher-order effect, which is called flexion, is described by two
parameters.  The spin-1 flexion is the complex derivative of $\kappa$
\be
{\cal F}= \nabla_{\rm c} \kappa\equiv {\dc \kappa \over \dc\theta_1}
+ {\rm i}{\dc\kappa \over \dc\theta_2}\; ,
\ee
and the spin-3 flexion is the complex derivative of $\gamma$,
\be
{\cal G}= \nabla_{\rm c} \gamma\;.
\ee

\subsection{Radial and tangential flexion}

The spin-1 flexion is a vector-like quantity, the gradient of the
surface mass density. Therefore, the spin-1 flexion is directed
towards the center of the lens in the case of an axi-symmetric mass
distribution. However, if the lens deviates from axial symmetry, e.g.,
due to elliptical halos or mass substructure
\citep{2009arXiv0909.5133B,2009MNRAS.400.1132H}, the flexion vector
will have a different direction.

In galaxy-galaxy lensing, shear can be decomposed into tangential and
cross components. Analogously, we decompose the spin-1 flexion into a
radial and tangential flexion component. They are defined as
\bea
{\cal F}_R &\equiv& - \vc{\cal F} \cdot \hat r \; ;
\elabel{rflexion}\\
{\cal F}_T &\equiv& - \vc{\cal F} \cdot \hat\phi \; ,
\elabel{tflexion}
\eea
where the spin-1 flexion vector is given by $\vc{\cal F}=({\cal F}_1, {\cal
  F}_2)=\nabla\kappa$ in Cartesian coordinates. $\hat r$ and $\hat
  \phi$ are the unit 
direction vectors in polar coordinates. 
For the spin-3 flexion, there is no such a clear intuitive picture of its
two components; thus, we only consider the spin-1 flexion in this
paper.

\begin{figure}
  \centerline{
    \includegraphics[width=6cm,height=6cm]{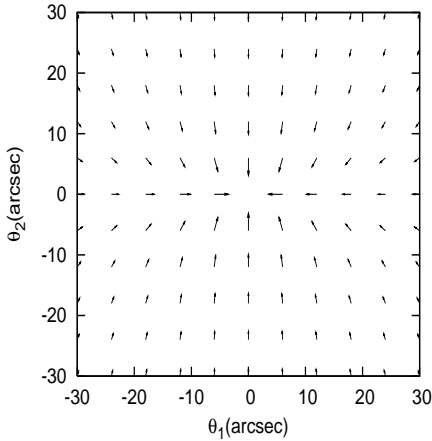}}
  \caption{
    Spin-1 flexion vector field for an elliptical isothermal density
    distribution with $\epsilon=0.3$. $\vc{\cal F}$ only points towards the
    center when the background galaxy is located on the major or minor axis of
    the halo. Note that the length of the vectors are logarithmically
    scaled, for better visibility
  }
  \label{fig:f1vec}
\end{figure}

As mentioned before, the spin-1 flexion vector is not directed towards the
center of an elliptical lens (Fig.\ts\ref{fig:f1vec}). Therefore, the
tangential flexion no longer vanishes and can be used to estimate the
ellipticity of the mass distribution.

\subsection{Elliptical mass distributions}

We now assume that the isodensity contours of the mass distribution are
ellipses with ellipticity $\eps$, or equivalently, axis ratio
$(1-\eps)/(1+\eps)$, and orientation $\phi_0$ of the major axis. In
this case, the surface mass density can be written as
\be
\kappa(\vc\theta)=K\rund{\theta\sqrt{{\cos^2(\phi-\phi_0)\over
      (1+\eps)^2} +{\sin^2(\phi-\phi_0)\over (1-\eps)^2}}}\;,
\elabel{kapp}
\ee
where the function $K(\rho)$ describes the radial density profile. The
derivatives of $\kappa$ with respect to radial and azimuthal
coordinates are given as
\bea
{\partial\kappa\over\partial\theta}&=&K'\;\sqrt{{\cos^2(\phi-\phi_0)\over
      (1+\eps)^2} +{\sin^2(\phi-\phi_0)\over (1-\eps)^2}}=-{\cal F}_R\;, \\
{1\over\theta}{\partial\kappa\over\partial \phi}&=&
{K'\;2\eps(1-\eps^2)^{-2}\sin2(\phi-\phi_0)\over
  \sqrt{{\cos^2(\phi-\phi_0)\over 
      (1+\eps)^2} +{\sin^2(\phi-\phi_0)\over (1-\eps)^2}}}=-{\cal F}_T\;.
\eea
Both of these derivates have the same radial profile, given by
$K'$. We now define the {\it flexion ratio} as
\be
r\equiv \abs{{\cal F}_T \over {\cal F}_R}\;,
\elabel{fratio}
\ee
which is the tangent of the angle between the direction to
the mass center and the direction of the flexion vector. For an
elliptical mass distribution, this becomes
\be
r(\phi)={|2 \epsilon \sin 2(\phi-\phi_0)| \over 1-2\epsilon\cos2(\phi-\phi_0)
  +\epsilon^2}\;.
\elabel{ratio-phi}
\ee
Thus, the flexion ratio $r(\phi)$ depends only on the ellipticity
$\eps$ and the orientation $\phi_0$ of the mass distribution, not on
its radial profile. In particular, it is independent of the lens
strength, and thus of the source and lens redshift.
In principle, from the measurement of the flexion
ratio at two polar positions $\phi$, one can determine both $\eps$ and
$\phi_0$, though due to noise, this determination will have large
uncertainty. Alternatively, one can consider the flexion ratio
averaged over all polar angles,
\bea
\langle r\rangle &=& {1\over 2\pi} \int_0^{2\pi} \d \phi
\abs{\frac{2\epsilon\sin 2(\phi-\phi_0)}{1- 2\epsilon \cos 2(\phi-\phi_0)
    +\epsilon^2}} \nonumber\\
&=&{2\over \pi}\, {\rm ln}{1+ \epsilon  \over 1-\epsilon}\;,
\elabel{rmean}
\eea
which depends solely on the ellipticity $\epsilon$. Due to its
simplicity, this mean flexion ratio can be easily 
measured from a flexion field in a given aperture.

However, we have to assume to know the center of the lens for
calculating the two flexion components. For a galaxy, the center is
assumed to be the bright center of galaxy. For clusters, the location
of the BCG not necessarily coincides with the mass center.  Moreover,
`intrinsic flexion' also introduces extra noise (see below).

For a given value of $\epsilon$, the flexion ratio is bounded. One can
see from (\ref{eq:ratio-phi}) that
\be
r(\phi) \leq r_{\rm max} = {2\epsilon \over 1-\epsilon^2}.
\elabel{rlimit}
\ee
Arbitrary large ellipticities, which would allow $r$ to become very
large, are implausible and most likely do not exist. We can take
limits on $\epsilon$ from other observations, e.g.,
\citet{2007ApJ...669...21P} compared galaxy-galaxy lensing along the
major and minor axes, and found that the axis ratio $b/a$ of galaxy
halos lies between 0.5 and 0.8 ($\epsilon \in [0.11,0.33]$). Moreover,
\citet{2010arXiv1004.4214O} analyzed 25 clusters and found
$\langle1-b/a\rangle =0.46$, corresponding to $\epsilon=0.32$.  Here
we use the conservative assumption that $\epsilon<0.8$, putting an
upper bound to the flexion ratio of $r<4.5$. We will employ this limit
to remove excessively large flexion ratios which are due to noise.

%

\section{Numerical test with NIE toy model}
In this section we describe some simulations which we have performed in order
to test the behavior of the estimator given in the previous section.
For this, we model the surface mass density profile by a non-singular
isothermal elliptical profile (NIE), described by
\be
K(\rho)={\theta_{\rm E}\over2\sqrt{\theta_{\rm c}^2+\rho^2}}\;,
\ee
where $\theta_{\rm E}$ is the Einstein radius, describing the strength
of the lens, and $\theta_{\rm c}$ is the core radius; for $\theta_{\rm
  c}=0$, this specializes to the singular isothermal elliptical
profile (SIE). In our simulations, 
we take 
$\theta_{\rm E}=6''$ and $\theta_{\rm c}=2''$. Galaxies are randomly
distributed within an $1'\times 1'$ 
`source plane' behind the lens. Resulting images are discarded if they
are located closer to the halo center than $6''$ (the strong lensing
regime) or at distances $|\vc\theta|>30''$, where the flexion signal
will be very small. Moreover, very large flexions cannot be measured
\citep{2008A&A...485..363S}, thus images with $|{\cal F}|>0.5$ are
discarded as well.

The flexion ratio is calculated according to
Eq.(\ref{eq:fratio}) for each point. A mean flexion ratio is obtained by
\be
\bar r= {1\over N} \sum_{i=1}^N r_i,
\elabel{ratiomean}
\ee
where $N$ is the number of images for each realization. Then the mass
ellipticity is estimated, according to Eq.(\ref{eq:rmean}), as
\be
\hat \epsilon= \dfrac{{\rm e}^{\pi \bar r /2}-1}{{\rm e}^{\pi \bar r /2}+1}.
\elabel{estimator}
\ee

\subsection{Noise-free case}

Due to the non-linearity, the estimator (\ref{eq:estimator}) is
expected to be biased. In order to test this, we generated mock
data sets with different $\epsilon=
0.03i$, $i=1,2,...20$. For each ellipticity, we used 20 realizations,
with a density of flexion points of $80\,{\rm arcmin}^{-2}$. The
filtering described above leads to 63
flexion data for each realization on average.

\begin{figure}
  \centerline{
    \includegraphics[width=7cm,height=7cm]{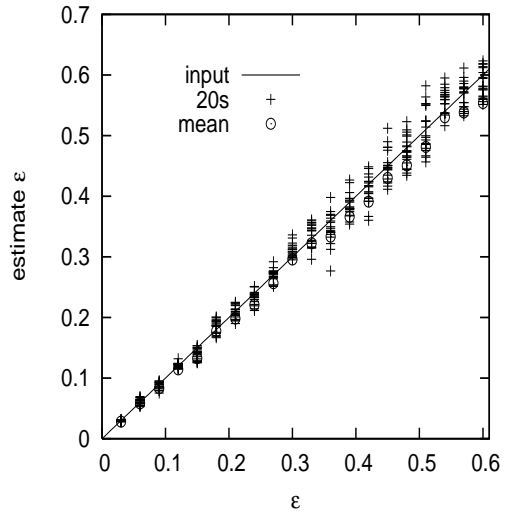}}
  \caption{
    Comparison of the ellipticity estimator (\ref{eq:estimator}) with the
    input ellipticity (solid line). The estimated ellipticity for 20
    realizations each for
    20 input values of $\epsilon$ are shown by pluses, for NIE models
    with $\theta_{\rm c}=2''$
  }
  \label{fig:rmean}
\end{figure}

For each realization, we calculated $\bar r$ from
Eq.(\ref{eq:ratiomean}) and estimate $\epsilon$ using
Eq.(\ref{eq:estimator}). In Fig.\ts\ref{fig:rmean}, we show
$\hat\epsilon$ vs. the input value of $\epsilon$. The solid line shows
the identity, the plus points are the estimates from the 20
realizations. In fact, the estimates $\hat\epsilon$ closely trace the
input value. The variance between different realizations becomes
larger with increasing $\epsilon$. The mean over the 20
realizations is shown by the plus points in
Fig.\ts\ref{fig:int-noise}, showing that
$\hat\epsilon$ very slightly underestimates the input value.


%
%
\subsection{Intrinsic flexion}
Our main source of noise is intrinsic flexion, meaning that the
sources can have non-vanishing third-order brightness
moments. Depending on the method of flexion measurement, the intrinsic
noise may be different \citep{2007ApJ...660.1003G}. Since flexion has
the dimension of an inverse length, the intrinsic flexion is also
inversely proportional to the image size. Therefore, the distribution
of intrinsic flexion depends on the survey and is difficult to obtain
from real measurements on current data. Effects of a
point spread function (PSF) need to be considered, which is far more
difficult for flexion than for shear measurements. In particular, an
anisotropic PSF may affect the direction of the flexion
vector.

We use a simple model to generate intrinsic noise for our simulated
data, by setting ${\cal F}^{\rm
  obs}={\cal F}_1+n_{f1} + {\rm i}({\cal F}_2+n_{f2})$. The components
of the intrinsic flexion
$n_{f1},n_{f2}$ are drawn from a Gaussian distribution, with
each component being characterized by $\sigma_{{\cal
    F}1}=\sigma_{{\cal F}2}=0.03$ arcsec$^{-1}$.

\begin{figure}
  \centerline{
    \includegraphics[width=7cm,height=7cm]{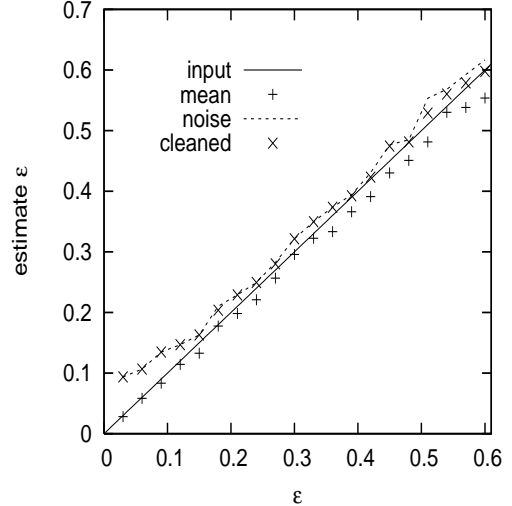}}
  \caption{
    Comparison of the inputting ellipticity (solid line) and the
    estimate (similar as Fig.\ref{fig:rmean}). The plus points are average of
    estimate for 20 realizations without noise. The other are the result from
    data with intrinsic noise. All the result are calculated from
    Eq.(\ref{eq:estimator}). The cross points are calculated while the points
    which give flexion ratio higher than the up limit (Eq.\ref{eq:rlimit})
    are excluded.
  }
  \label{fig:int-noise}
\end{figure}

In Fig.\ts\ref{fig:int-noise} we compare the estimates $\hat\epsilon$
with the input values (solid line). The plus points show the
mean over 20 realizations without noise (Fig.\ts\ref{fig:rmean}), and
the dashed line displays estimates with intrinsic noise included. Not
surprisingly, our estimates from noisy simulations are larger than the
input values (intrinsic noise will cause the estimate to deviate from
zero even for a perfectly symmetric mass distribution).  This is
particularly significant for small $\epsilon$.  As an additional
test, we reject flexion data which have $r>4.5$, and perform our
estimate again. The result is shown by crosses in
Fig.\ts\ref{fig:int-noise}. One can see that the bias is slightly
reduced in this case, in particular for larger $\epsilon$, but still
$\hat\epsilon$ overestimates the true ellipticity. The amplitude of
this bias depends of course on the intrinsic flexion distribution.

\begin{figure*}
  \centerline{
    \includegraphics[width=7cm,height=7cm]{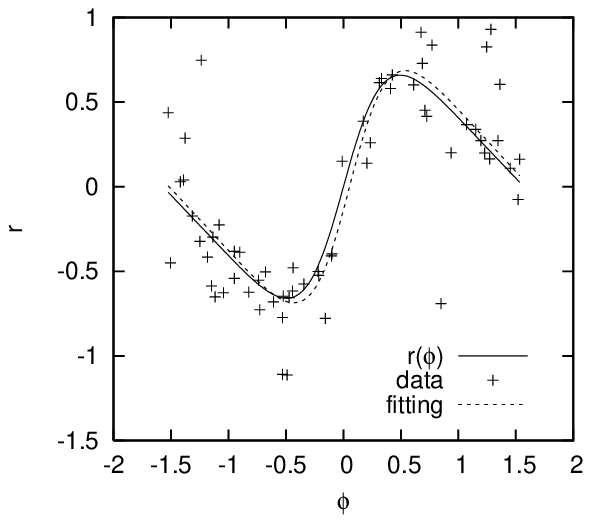}
    \includegraphics[width=7cm,height=7cm]{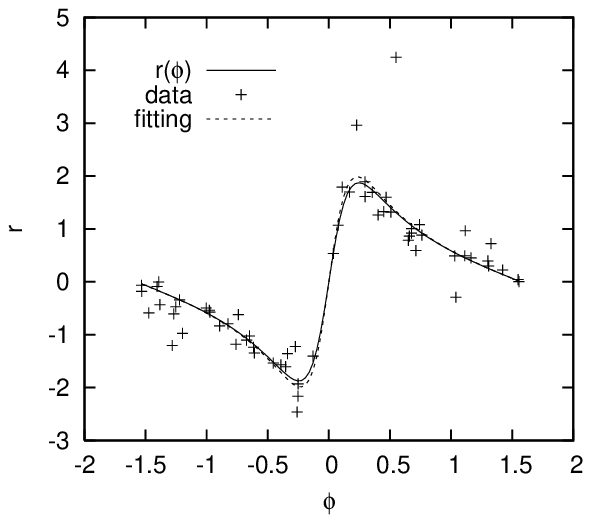}}
  \caption{
    The flexion ratio $r$ varies with $\phi$. The dash line is for an SIE halo
    model. The points are calculated from the simulated data,
    the dotted line is the fitting result with Eq.(\ref{eq:ratio-phi}).
    Left is halo with $\epsilon=0.3$ and right is $\epsilon=0.6$.
  }
  \label{fig:rphifit}
\end{figure*}

Another way to estimate the ellipticity of the mass distribution is to
fit the flexion ratio with the model (\ref{eq:ratio-phi}),
with the two
free parameters $\epsilon$ and $\phi_0$. The
result for two realizations are presented in
Fig.\ts\ref{fig:rphifit}. The points are calculated from simulated
data of an NIE with $\epsilon=0.3$ (left) and $\epsilon=0.6$
(right). The dotted lines are the fitting result with
Eq.(\ref{eq:ratio-phi}) whereas the solid curves 
show Eq.(\ref{eq:ratio-phi}) with the input ellipticity.
The fits almost perfectly agree with the input
model. For the left panel, the fitting yields $\epsilon=0.31$ and
$\phi_0=0.1$, whereas 
$\hat\epsilon=0.36$. In the right panel, $\epsilon=0.62$ and
$\phi_0=0.006$ from fitting, whereas $\hat\epsilon=0.67$.  In both panels of
Fig.\ts\ref{fig:rphifit}, there are several points with significant
deviations from the fitting curves which is the reason for
$\hat\epsilon$ to be larger than the input value, whereas these
points do not strongly affect the fitting result.

Obviously, the fitting method yields a more accurate estimate of the
mass ellipticity than the estimator $\hat\eps$, and at the same time
also estimates the orientation. It is therefore the preferred method
for individual lenses. In contrast, the estimator $\hat\eps$ can be
applied also to an ensemble of lenses, by superposing their respective
flexion values. In this way, $\hat\eps$ estimates the weighted mean
ellipticity of the ensemble of lenses. Note that in contrast to
galaxy-galaxy lensing with the shear method, the individual lenses do
not have to be aligned before the averaging; hence, no assumption
about the relative orientation of mass and light needs to be made.

\subsection{Centroid offset test}
\begin{figure}
    \includegraphics[width=7cm,height=6cm]{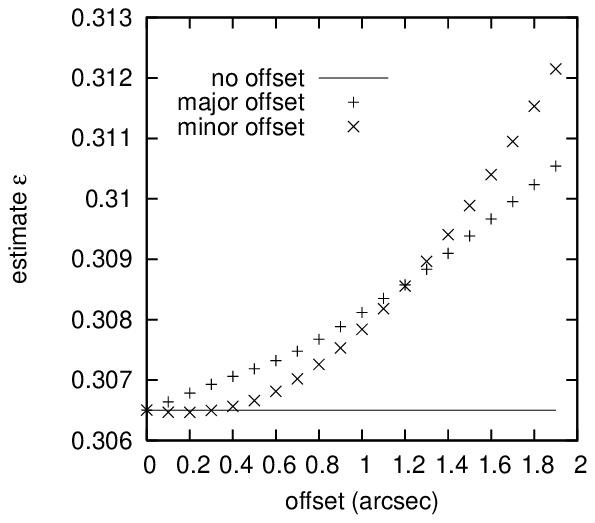}\\
    \includegraphics[width=7cm,height=6cm]{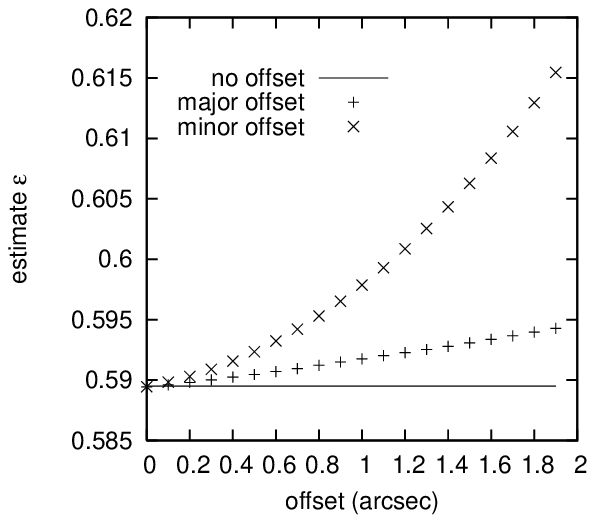}
    \caption{ Bias due to the centroid offset. The
      solid line is our estimate (400 sets of data) without
      centroid offset. The 
      plus (cross) points are estimate for the centroid offset alone
      the major (minor) axis of the elliptical halo.  The top (bottom)
      panel is for galaxy halo with ellipticity 0.3 (0.6).  }
  \label{fig:ernoise}
\end{figure}

Since the ellipticity represents the deviation from symmetry of a mass
distribution, a centroid offset will affect the determination of the
ellipticity, leading to an additional bias. We simulated this effect,
by calculating the radial and tangential flexion components with
respect to $\vc\theta_0+\delta\vc\theta$, where $\vc\theta_0$ is the
true center of the lens, and $\delta\vc\theta$ is the offset. For
simplicity, we choose two sets of $\delta\vc\theta$, one along the
major axis, the other along the minor axis of the mass
distribution. To isolate this effect, we generate 400 data sets without
intrinsic noise, for two NIE with $\epsilon=0.3,0.6$. The results are
shown in Fig.\ts\ref{fig:ernoise}. A centroid offset indeed biases the
estimate of the ellipticity, but the magnitude of this effect is
relatively small, as long as the offset is considerably smaller than
the Einstein radius of the lens. For the lens with higher ellipticity
(bottom panel), an offset along the minor axis has a larger impact on
the resulting ellipticity than one along the major axis. We conclude
that the effect of misidentifying the lens centroid is of little
concern for galaxy lenses, but may be more important for galaxy
clusters where the centroid is often ill-defined from observations.

\section{Conclusions and Outlook}

In this paper, we have studied galaxy-galaxy flexion with elliptical
mass distributions. We derived the ratio $r$ of radial and tangential
flexion and found that it is independent of the radial mass profile of
the lens, and of the source and lens redshift. The flexion ratio
depends solely on the ellipticity and orientation of the mass
distribution. We defined an estimator $\hat\eps$ for the ellipticity in
terms of the flexion ratio, and tested its performance with simple
simulations. This estimator does not rely on knowing the orientation
of the distribution, i.e., assuming that it follows the orientation of
the light; hence, it can be used to statistically superpose several
lenses and thereby obtain their average ellipticity. The independence
of this estimator from an assumed orientation is quite different from
methods to constrain lens ellipticities from shear
\citep{2000astro.ph..6281B,2004ApJ...606...67H}, for which the
orientation needs to be known. In the appendix we show that an
analogous method to that presented here does not work for shear, but
is unique to the spin-1 nature of the first flexion.

For an individual lens with several flexion measurements, the
ellipticity and orientation can be derived by fitting the measured
flexion ratios with these two parameters. The accuracy of the
resulting estimate is substantially better than that of $\hat\eps$.

Our simulations showed that $\hat\eps$ is biased. A first bias is due
to the non-linearity of the relation between estimated ellipticity and
mean flexion ratio. The size of this bias depends on the number
density of sources for which flexion can be measured, and disappears
in the limit of very high source density (or large number of lenses
for which the mean ellipticity is explored).  Intrinsic flexion causes
a more significant bias, in particular for small values of
$\epsilon$. A misidentification of the mass centroid is another source
of bias; however, if the centroid offset is much smaller than the
Einstein radius of the lens, this bias is rather small.

There will be several complications in a real analysis. First, the
flexion ${\cal F}$ can not be directly measured, but only reduced
flexion \citep{2008A&A...485..363S}.  
This only requires a small
modification, though: if $\kappa$ is constant on ellipses, so will be
$\ln(1-\kappa)$. Since
\be
\nabla_{\rm c}\ln (1-\kappa)={G_1-g G_1^*\over 1-g g^*}\;,
\elabel{redflex}
\ee
where $G_1$ is the spin-1 reduced flexion, $G_1=\nabla^*_{\rm c}
g=({\cal F}+g {\cal F}^*)/(1-\kappa)$, and $g=\gamma/(1-\kappa)$ the
reduced shear, we can use the phase of (\ref{eq:redflex}) instead of
the flexion ratio. Furthermore, the impact of
intrinsic flexion is uncertain, reflecting our lack of knowledge about
its magnitude which is ill-determined from current observations.
A further careful study of intrinsic flexion will be of interest before the
flexion method is applied to real data. 

In addition, as we have shown, a low number density of background sources also
introduce bias. Thus a deep survey with more background sources is
clearly favoured.
Current surveys such as the Hubble Space Telescope (HST), the
Canada-France-Hawaii Telescope Legacy Survey (CFHTLS) or Subaru telescope
might be able to constrain halo ellipticity using flexion, if the flexion can
be measured with sufficient accuracy. The James Webb Space Telescope
will almost certainly allow very accurate measurements of the
ellipticity of galaxy- and group-sized lenses.

\section*{Acknowledgments}
We thank Dave Goldberg and Ismael Tereno for very useful comments on
the manuscript.  This project is supported by the Deutsche
Forschungsgemeinschaft under the project SCHN 342/7.

\begin{appendix}
\section{\label{sc:A}The shear profile in SIE halo}
In this appendix, we present the tangential and cross shear profile
for an SIE halo model and show that one cannot apply the ratio of
cross to tangential shear to estimate the ellipticity of the lens.

To calculate the shear, we calculate the deflection potential using the
Poisson equation
\be
{1\over \theta}{\dc\over\dc \theta}\rund{\theta {\dc \psi\over \dc \theta}}
+{1\over \theta^2}{\dc^2\psi \over\dc \phi^2}=2\kappa\;.
\ee
The convergence can be written in polar coordinates
\be
\kappa(\theta,\phi)={\theta_{\rm E}(1-\epsilon)\over 2\theta} \dfrac{1}
{\sqrt{f^2\cos^2\phi +\sin^2\phi}},
\elabel{siepolar}
\ee
where $f$ is the axs ratio. The solution of the Poisson equation can
be found in 
\citet{1994A&A...284..285K}, but here we write it in our notation as
\bea
\psi(\theta,\phi)\!\! &=&\!\! \dfrac{\theta_{\rm E}(1-\epsilon) \theta}{f^-}\\
&\times&\!\! \eck{\arcsin
\rund{f^-\cos \phi} \cos\phi +{\rm arcsh}\rund{{f^-\over f}
  \sin\phi}\sin\phi},\nonumber
\eea
where $f^-=\sqrt{1-f^2}$. Then we rewrite the potential in Cartesian
coordinates and calculate the shear by $\gamma_1=(\psi_{,11}-\psi_{,22})/2$ and
$\gamma_1=\psi_{,12}$. We thus obtain the two shear component for the
SIE halo, 
\bea
\gamma_1 &=&-\dfrac{\theta_{\rm E} (1-\epsilon)(\theta_1^2-\theta_2^2)}
{2\theta^2 d}\;; \\
\gamma_2 &=&-\dfrac{ \theta_{\rm E} (1-\epsilon) \theta_1\theta_2}
{\theta^2 d}\;,
\eea
where $d=\sqrt{f^2\theta_1^2 +\theta_2^2}$. According to the definition
$\gamma_{\rm t} = -\gamma_1\cos2\phi -\gamma_2\sin2\phi$, 
$\gamma_{\times} = \gamma_1\sin2\phi -\gamma_2\cos2\phi$,
the tangential-to-cross shear for SIE halo reads
\bea
\gamma_{\rm t} &=& {\theta_{\rm E}(1-\epsilon) \over 2 d};\\
\gamma_{\times} &=& 0.
\eea

\end{appendix}
\bibliographystyle{aa}
\bibliography{/users/xer/bib/refbooks,/users/xer/bib/reflens,/users/xer/bib/flexion,/users/xer/bib/refcos,/users/xer/bib/simcos}

\end{document}